\documentclass{article}
\usepackage{fixltx2e}
\usepackage[latin1]{inputenc}
\usepackage{amsmath,amsfonts,amssymb,bm}
\usepackage{graphicx}
\usepackage{cite}
\usepackage[tight,hang,small,bf]{subfigure}

\begin{document}

\title{Fast optical source for quantum key distribution based on semiconductor optical amplifiers}

\author{M.~Jofre,~A.~Gardelein,~G.~Anzolin,~W.~Amaya,\\~J.~Capmany,~R.~Ursin,~L.~Peñate,~D.~Lopez,\\~J.L.~San~Juan,~J.A.~Carrasco,~F.~Garcia,~F.J.~Torcal-Milla,\\~L.M.~Sanchez-Brea,~E.~Bernabeu,~J.M.~Perdigues,~T.~Jennewein,\\~J.P.~Torres,~M.W.~Mitchell~and~V.~Pruneri}

%\address{$^{1}$ICFO-Institut de Ciencies Fotoniques, Mediterranean Technology Park, 08860 Castelldefels (Barcelona), Spain\\
%$^{2}$Institute of Telecommunications and Multimedia Applications (ITEAM), Universidad Politécnica de Valencia, Spain\\
%$^{3}$Institute for Quantum Optics and Quantum Information (IQOQI), Austrian Academy of Sciences, Austria\\
%$^{4}$Faculty of Physics, University of Vienna, Austria\\
%$^{5}$ALTER Technology Group, Spain\\
%$^{6}$LIDAX, 28850 Torrejón de Ardoz, Spain\\
%$^{7}$EMXYS, 03206 Elche, Spain\\
%$^{8}$Applied Optics Complutense Group (AOCG), Universidad Complutense de Madrid, Spain\\
%$^{9}$European Space Agency, Noordwijk, The Netherlands\\
%$^{10}$Institute for Quantum Computing, University of Waterloo, Canada\\
%$^{11}$Dept. of Signal Theory and Communications, Universitat Politècnica de Catalunya, Spain\\
%$^{12}$ICREA-Institució Catalana de Recerca i Estudis Avançats, 08010 Barcelona, Spain}
%
%\email{marc.jofre@icfo.es} %% email address is required

\date{}



\maketitle

\begin{abstract}
A novel integrated optical source capable of emitting faint pulses
with different polarization states and with different
intensity levels at $100$ MHz has been developed. The source relies
on a single laser diode followed by four semiconductor optical amplifiers
and thin film polarizers, connected through a fiber network. The
use of a single laser ensures high level of indistinguishability in
time and spectrum of the pulses for the four different
polarizations and three different levels of intensity. The
applicability of the source is demonstrated in the lab through a
free space quantum key distribution experiment which makes
use of the decoy state BB84 protocol. We achieved a lower bound secure key rate of the order
of $3.64$ Mbps and a quantum bit error ratio as low as $1.14\times
10^{-2}$ while the lower bound secure key rate became $187$ bps for an equivalent attenuation of $35$ dB. To our knowledge, this is the fastest polarization encoded QKD system which has been reported so far. The performance, reduced size, low power consumption and
the fact that the components used can be space qualified make the
source particularly suitable for secure satellite
communication.
\end{abstract}

%%%%%%%%%%%%%%%%%%%%%%%%%%  body  %%%%%%%%%%%%%%%%%%%%%%%%%%
\section{Introduction}
\textit{Quantum key Distribution} (QKD), guarantees absolutely
secure key distribution, since making use of the principles
of quantum physics \cite{Bennett1984}, using the fact that it is not possible to measure or copy an unknown quantum state without being detected \cite{RevModPhys.77.1225}.
A cryptographic key is generated out from the measurement of the
information encoded into specific quantum states of a photon (e.g.
polarization or phase). The first QKD scheme,
proposed by Bennett and Brassard \cite{Bennett1992}, employed
single photons sent through a quantum channel, plus classical
communications over a public channel to generate a secure shared
key. This scheme is commonly known as the BB84 protocol.
Attenuated laser pulses or \textit{faint pulse sources} (FPS),
which in average emit less than one photon per pulse, are often
used as signals in practical QKD devices. With the introduction of
the decoy state protocol \cite{Hwang2003,Wang2005a,PhysRevLett.94.230504}, faint pulse
systems can offer comparable QKD security with respect to single
photon sources, in particular for high loss situations \cite{Nauerth2009}.

Current photon-detector technology and optical guiding media
losses limit QKD on Earth to $250$ km in optical fibers
\cite{Stucki2009} and $144$ km in free-space links
\cite{Schmitt-Manderbach2007}. The unique features of space
can potentially offer extremely long propagation paths, essential
for the realization of a global QKD network
\cite{PerdiguesArmengol2008165,ursin-2008}. Nevertheless,
many components of the system, besides the source, require special
attention for the implementation of QKD. The development of a
source suitable for space, apart from the optical design, requires
a demanding opto-mechanical engineering as well as high level of integration
with the electronics.

In this paper, we report the development of a novel, compact and
reliable FPS-based \textit{semiconductor optical amplifiers}
(SOAs) which emits pulses at a repetition
rate of $100$ MHz, suitable for free-space QKD applications. It
has been measured that the source achieves a Lower Bound Secure Key Rate of $3.64$ Mbps implementing the decoy state BB84
protocol. The source is capable of generating pulses with at least three different intensity levels (i.e.
number of photons per pulse) and four different polarization
states. The demonstrated FPS ensures high level of
indistinguishability among the different intensity and
polarization pulses and ensures phase incoherence of consecutive
generated states. It is based on a single diode emitting a
continuous optical train of pulses externally modulated in
intensity and polarization by using a combination of four SOAs and
polarizing optical elements. The wavelength, reduced power
consumption, compactness and space qualifiable optoelectronic
components constituting the source make it very suitable for space-based
QKD transmission. Although the proposed source has been conceived
for free space QKD at around $850$ nm, the concept can be extended
to other wavelengths (e.g. $1550$ nm) and other media, including
optical fibers.

\section{The compact faint pulse source}\label{FPSSetup}
\begin{figure}[htbp]
\centering\includegraphics[width=13cm]{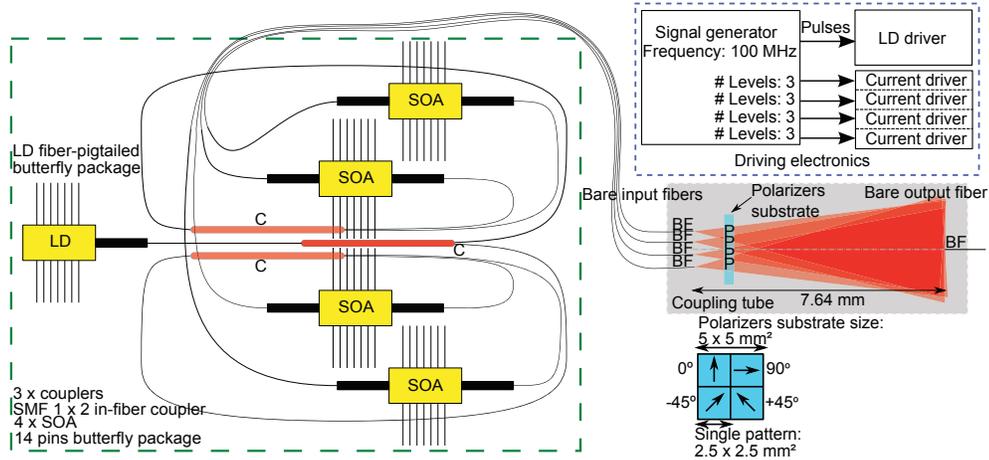}
\caption{Schematic of the QKD source. [LD] denotes a laser diode,
[C] a single-mode in-fiber $1x2$ coupler, [SOA] a semiconductor
optical amplifier, [P] a free-space thin-film polarizer and [BF] a
single-mode bare fiber.} \label{Fig:FPS_4SOA_Source}
\end{figure}

In order to use it for space applications, the proposed integrated
FPS source consists of space-qualified discrete components (Figure
\ref{Fig:FPS_4SOA_Source}): a single semiconductor laser diode
emits a continuous train of optical pulses at $100$ MHz,
equally split to four outputs using three in-fiber single-mode
$1x2$ couplers. Each fiber output is followed by an integrated
(waveguide) SOA. The four bare fibers to the coupling tube are
accurately positioned on a custom opto-mechanical mount to
simultaneously achieve the correct launching to the polarizers
substrate sheet as well as to introduce a desired $70$ dB
coupling-loss to work in the single-photon regime at the end of
the output bare fiber. Each polarizer is accurately oriented to produce one of the four polarizations ($0$\textdegree, $90$\textdegree,$45$\textdegree and
$-45$\textdegree) required for BB84 protocol. Finally, the output fiber can be connected
to the corresponding optical link interface (e.g. a telescope for
free-space communication or optical fiber for terrestrial links).

The \textit{distributed feedback} (DFB) \textit{laser diode} (LD)
source is directly modulated at $100$ MHz by a train of electrical pulses of
about $1$ ns duration. The generated optical pulses do not have
any phase coherence among them due to the fact that the laser is
set below and above threshold from one pulse to the subsequent
one, thus producing a random phase for each pulse
\cite{Jofre2010JLT-FPS-IMPM}. In this way phase coherence of consecutive generated states is absent, which otherwise would be
detrimental for the link security. Single-mode $1x2$ in-fiber
couplers perform the task of passively splitting the continuous optical pulse train into four equal outputs, while
sufficiently preserving the polarization state linearity generated
by the LD, before entering the SOAs. 

For the implementation of a
QKD system using decoy state protocol, besides four different
polarization states ($0$\textdegree,
$90$\textdegree,$45$\textdegree and $-45$\textdegree), the FPS
source should generate three intensity levels (optimally $1/2$, $1/8$
and $0$ photons in average per pulse \cite{Wang2005b,PhysRevA.72.012326}) in
order to operate in the single photon regime while optimizing the
decoy state protocol. Each SOA performs the double task of spatial
switching and amplitude modulation; it selects a specific fiber
output while, by changing the driving voltage, it generates the
three intensity levels needed for the decoy state protocol. Note that the spatial switching among the four fiber outputs is then transformed in polarization modulation, according to the scheme described above, by going through different polarizing patterns of the polarizers substrate. In this way intensity and polarization modulation can be achieved with high \textit{extinction ratio} (ER), without the need of using \textit{Polarization Maintaining Fibers} (PMFs) and PMF couplers or the complexity of having to maintain a specific input polarization state or high \textit{Degree of Polarization} (DOP) along the whole optical assembly. Usually, the common
parameter of interest for a SOA is the gain. However in this case
the frequency response to an electrical signal is equally
important, in particular for the time and spectral
indistinguishability of the pulses with different polarizations
and energy levels.

A single LD outperforms, as commented in Section \ref{Sec:Source_Security_Analysis}, the achievable security of the qubits sent by the source compared to using more than one LD, while using a DFB LD generates short optical pulses close to transform limited pulses which ensures the correct functionality and minimum additional distortion of any subsequent optical devices and components. The time duration and spectral bandwidth product of the generated pulses is $0.56$, very close to the theoretical value for transform limited Gaussian pulses ($0.44$). The main reason of using SOAs as active switching and intensity selection devices is their high ER ($>20$ dB) which is highly desirable when implementing 3-state decoy protocol. SOAs outperform other solutions such as Mach-Zehnder or Electro-absorptive modulators which have lower ER when operated with optical pulses as input signal. Finally, thin-film polarizers achieve high DOP of the order of $99.68$\%, being a remarkable specification in order to ensure an adequate quantum bit error ratio and thus high efficiency of secure key generation rate. Polarization modulators can not achieve such high values of DOP.

\section{Experimental measurements}\label{Experimental_measurements}
\begin{figure}[htbp]
\centering
\includegraphics[angle=0,width=8cm]{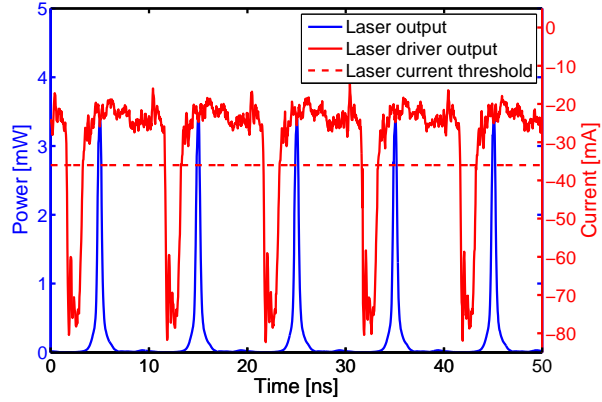}
\caption{Optical pulse train generation. Laser driver output (upper curve, right axis) produces few-ns, negative-going pulses which drive the DFB laser briefly above threshold (dashed-line, right axis). The laser output (lower curve, left axis), shows pulses with 100 MHz repetition rate and sub-ns duration.}
\label{Fig:Laser_output_and_Laser_driver_output_time}
\end{figure}

Figure \ref{Fig:Laser_output_and_Laser_driver_output_time} shows
the train of optical pulses generated by the laser diode when
driven by electrical pulses of $1$ ns at $100$ MHz. The resulting
optical pulse duration is about $400$ ps. Since all optical
pulses are generated in the same way, by direct modulation of the laser diode, they can be assumed to be
indistinguishable. The laser diode is driven in direct modulation with a strong RF driving pulse with $24$ mA DC bias current (below threshold, $36$ mA). Furthermore, the short optical pulse duration of
$400$ ps has the advantage of increasing
the signal-to-noise ratio, since the measurement window
(detection time) in the receiver side can small.

\begin{figure}[htbp]
\begin{center}
\subfigure[Drivers'
outputs.]{\label{SOAs_driver_outputs}\includegraphics[angle=0,width=0.49\columnwidth]{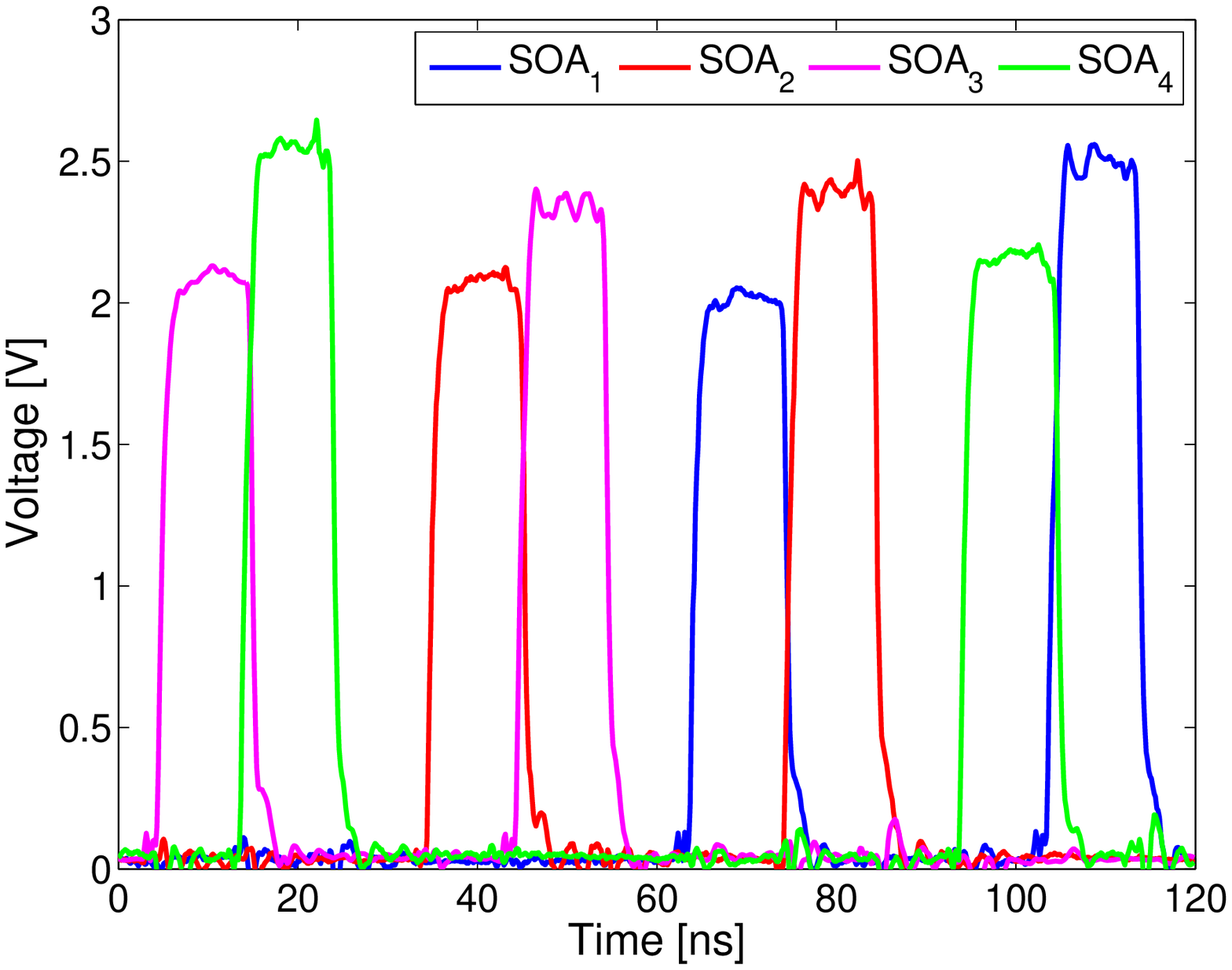}}
\subfigure[Optical
outputs.]{\label{SOAs_outputs}\includegraphics[angle=0,width=0.49\columnwidth]{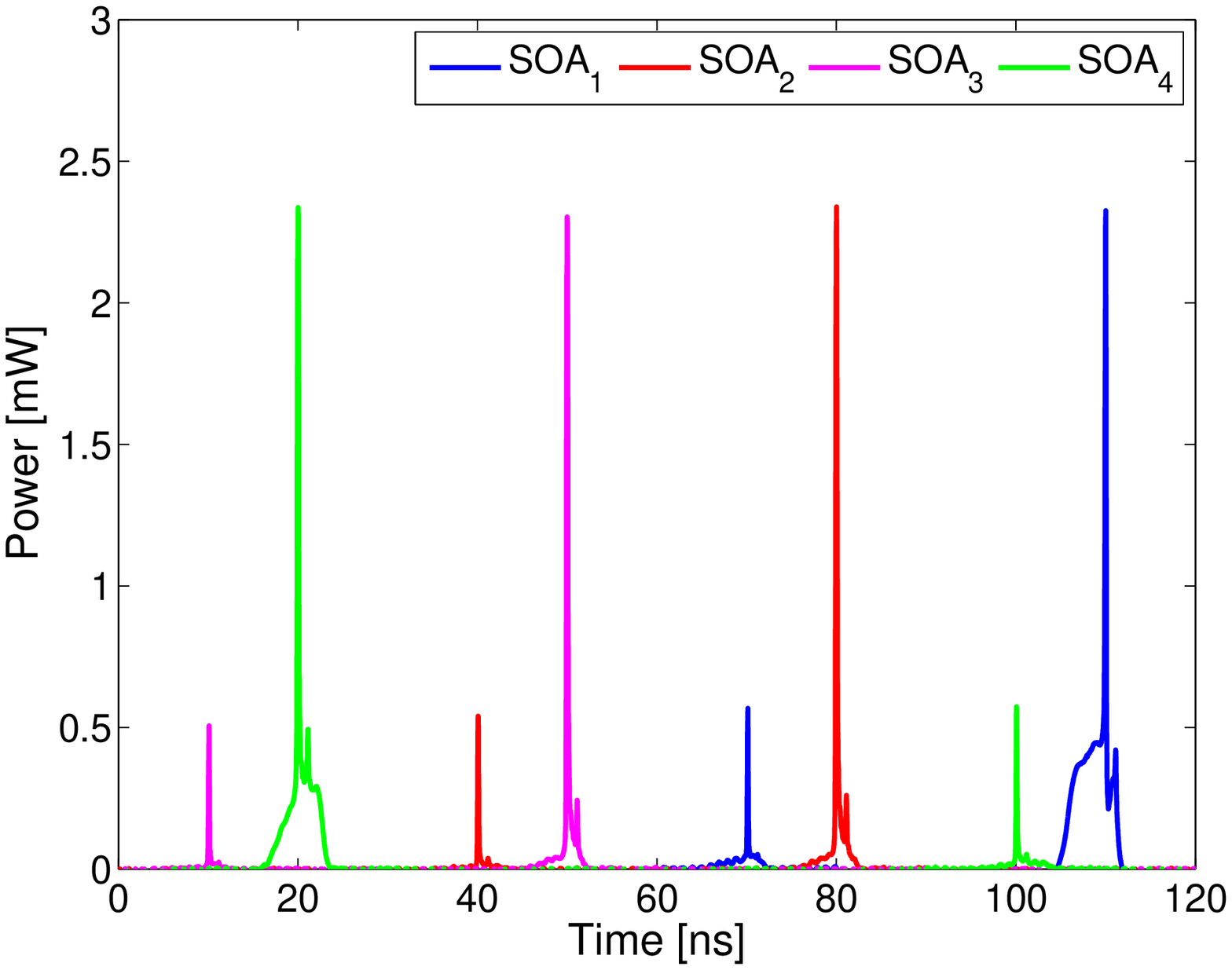}}
\caption{Driving signals to the four SOAs (a) and corresponding generated optical pulses (b), as required for the
implementation of the decoy state BB84 protocol. Note that the driving voltages are such that the peak optical intensities from the four SOAs are similar.}
\label{Fig:SOA_modulation}
\end{center}
\end{figure}

In Figure \ref{Fig:SOA_modulation}, the pulses with different
intensities required to implement the decoy state protocol are
shown for all the four SOAs, together with the driving signals,
loaded with a $50\Omega$ resistance for monitoring
purposes. Notice that the modulation time window of the driving signals is much larger ($>5$ ns) than the optical pulse thus minimum distortion is produced. A continuous periodic sequence, driving the four SOAs at
the same time, is used to set the intensity level of a single SOA
while the other three SOAs remaining in the OFF state. In
particular, the three intensity levels correspond to: a high
intensity level state which is set to $1/2$ average photon per
pulse by adjusting the coupling efficiency to the output bare
fiber; then a second lower intensity level and the third level is
set as vacuum (the SOA is not switched ON).

\begin{figure}[htbp]
\centering
\includegraphics[angle=0,width=8cm]{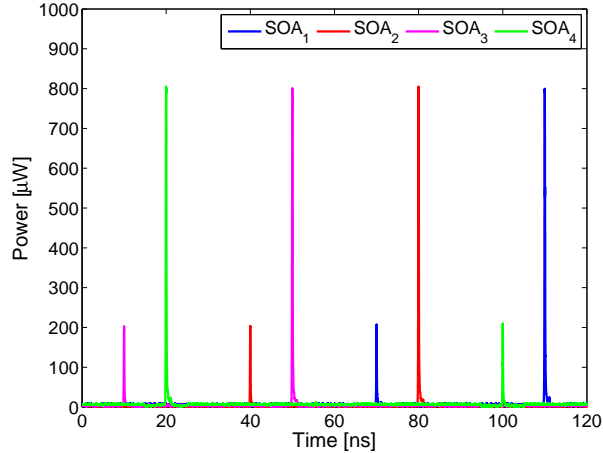}
\caption{Generated optical pulses after extensively removing ASE using a 3nm interference filter with central peak wavelength of 852nm.}
\label{Fig:SOAs_output_filtered}
\end{figure}

It is clear that the optical pulses from the SOAs have \textit{Amplified Spontaneous Emission} (ASE) noise reducing the indistinguishability of the different states, this can be mitigated by using proper filters, as shown in Figure \ref{Fig:SOAs_output_filtered}, and/or selection of SOAs. In fact, SOAs $2$ and $3$ produce nearly identical pulses without the use of any filter.

\section{Source security analysis}\label{Sec:Source_Security_Analysis}
\begin{figure}[htbp]
\begin{center}
\subfigure[Temporal profiles.]{\label{SOAs_temporal_filtered}\includegraphics[angle=0,width=0.5\columnwidth]{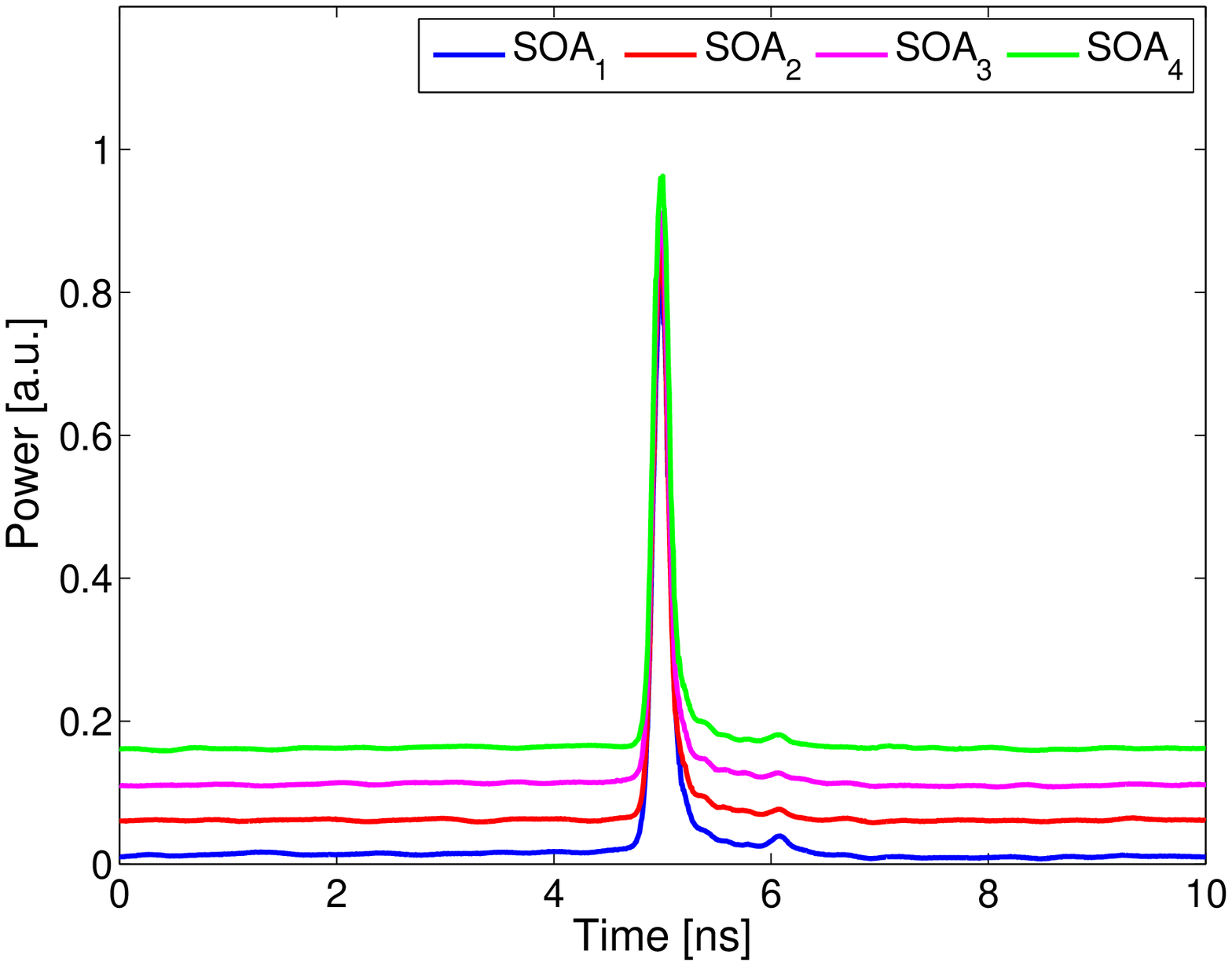}}
\subfigure[Spectral profiles.]{\label{SOAs_spectral_filtered}\includegraphics[angle=0,width=0.48\columnwidth]{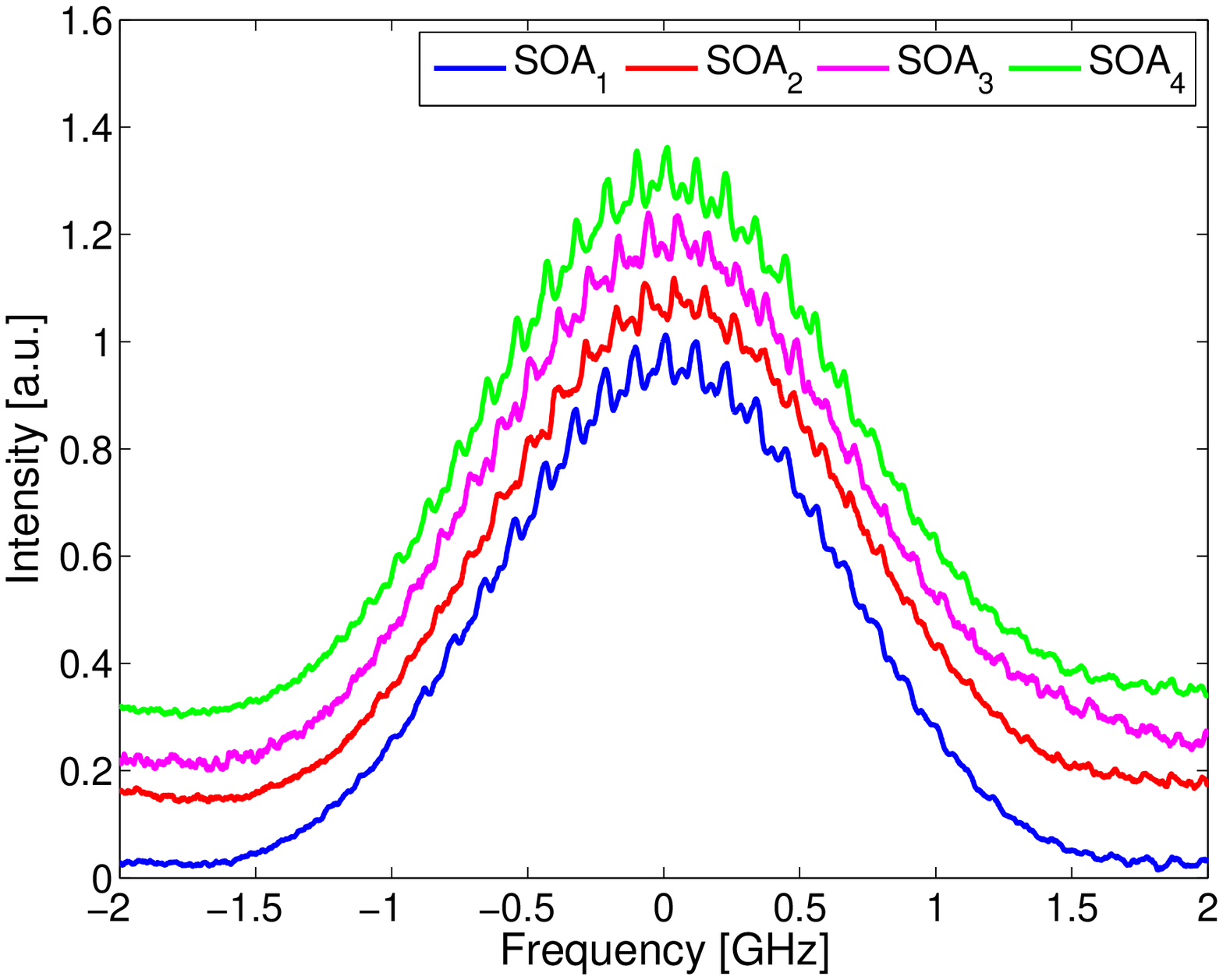}}
\caption{Temporal (a) and spectral (b) profiles for pulses with same intensity level (high) and four different polarization states, where an offset has been introduced in the vertical axis only for visualization purposes. As expected, these show a high degree of similarity, indicating minimal distortion of the pulses.}
\label{Fig:SOAs_temporal_spectral_indistinguishability}
\end{center}
\end{figure}

In all QKD implementations it is important to assess the actual security of the system, considering the imperfections of the different devices assembled. In this section the security of the presented source is computed with time and frequency resolution higher than the temporal and spectral shape widths of the generated optical pulses. Figure \ref{Fig:SOAs_temporal_spectral_indistinguishability} shows the temporal and spectral shapes of pulses with same intensity level and four different polarization states, filtered ASE, where an offset has been introduced in the vertical axis only for visualization purposes otherwise the pulses would highly overlap in intensity. A $8$ GHz amplified photodiode and a $4$ MHz resolution Fabry-Perot interferometer were used for the temporal and spectral measurements on indistinguishability, respectively, showing a high degree of similarity of the generated optical pulses.

Mutual information is the information shared by two random variables, $X$ and $Y$: it measures how much knowing one of these variables reduces the uncertainty on the other. For example, if $X$ and $Y$ are independent, then knowing $X$ does not give any information about $Y$ and vice versa, so their mutual information is zero. On the other extreme, if $X$ and $Y$ are identical then all information conveyed by $X$ is shared with $Y$: knowing $X$ determines the value of $Y$ and vice versa. Formally, the mutual information of two continuous random variables $X$ and $Y$ is defined as
\begin{equation}
I(X;Y)=\int_{Y}\int_{X}p\left(x,y\right)\log\left(\frac{p\left(x,y\right)}{p\left(x\right)p\left(y\right)}\right)dx\ dy
\end{equation}
where $I(X;Y)$ is the mutual information and $p\left(x,y\right)$ is the joint probability distribution between $X$ and $Y$ variables, $p\left(x\right)$ and $p\left(y\right)$ are the probability distributions of $X$ and $Y$ variables, respectively.

In particular, in FPS sources it is interesting to compute the average mutual information between the different bits sent $B$ and the physical parameters that could lead to side-channels, thus leaking information of the generated secure key to Eve if not accounted for. For the presented source, with filtered ASE, the mutual information for the relevant physical parameters with respect to the bit value $B$ are computed below:
\begin{itemize}
\item Spatial: The coupling output fiber also serves as a spatial filter, removing higher order modes. Because of the short length of the output SMF fiber (few centimeters) there might be still information transmitted by the fiber cladding. However this information leakage $I(S:B)$ is of the order of $10^{-5}$ bits per pulse.
\item Spectral: The mutual information between the bit value $B$ and the spectra $F$: $I(F:B)=1.75\times 10^{-3}$ bits per pulse.
\item Temporal: The mutual information between the bit value $B$ and the temporal shape $T$: $I(T:B)=1.92\times 10^{-3}$ bits per pulse, when removing the ASE. If the ASE is not removed, considering the generated optical pulses from the four particular SOAs available, the mutual information computes to $I(T:B)=7.25\times 10^{-2}$ bits per pulse, mainly because of high levels of ASE in SOAs $1$ and $4$. If SOAs were pre-selected to have low ASE the $I(T:B)$ could achieve the $10^{-3}$ bits per pulse order again.
\end{itemize}

Notice that in Figure \ref{SOAs_temporal_filtered} different pulses are superimposed achieving a good temporal overlap by careful matching the fiber's length with an accuracy of $1 mm$, not being necessary to further temporally adjust the overlap of the different optical pulses by using extra delay-adjustment hardware. Furthermore, it is clear that not perfect transform limited optical pulses are generated thus even similar optical pulses in their temporal and spectral envelopes could have distinguishing features in their phases. Using a single initial laser diode, considering that the distortion due to subsequent optical components is negligible (e.g. no significant chirp is introduced) because they are operated over a time scale much larger than the actual optical pulse duration, it ensures no distinguishing features in particular in the phases of the different optical pulses. While using initially different laser diodes, distinguishing features can be present in the phase of the different optical pulses which can not be avoided by proper filtering.

\section{QKD free-space transmission}
\begin{figure}[htbp]
\centering\includegraphics[width=12cm]{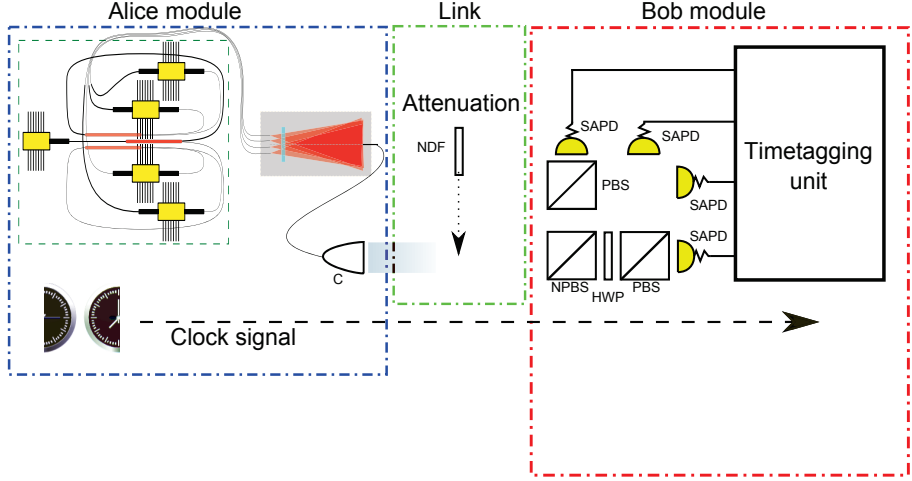}
\caption{Schematic of the QKD measurement setup. Alice module is
composed of the FPS also providing a common clock signal of $10$
MHz in order to synchronize Alice and Bob modules. [C] denotes a
fiber to free-space collimator, [NDF] a neutral density filter,
[NPBS] a non-polarizing beamsplitter, [HWP] a half-wave plate,
[PBS] a polarizing beamsplitter and [SAPD] a single-photon
detector.} \label{Fig:QKD_Alice_Bob_scheme}
\end{figure}

Figure \ref{Fig:QKD_Alice_Bob_scheme} shows the setup used
for the decoy state BB84 transmission measurement. The output bare
fiber was connected to a free-space collimator pointing to the polarization-sensitive detection module, a.k.a. ``Bob module'' commonly used for polarization-encoded QKD. The polarization detection fidelity, defined as the ratio between erroneous detected polarization states to total detected polarization states which defines the capability to correctly resolve the received polarization states, of the receiver module used in the experiment was $7.9\times 10^{-3}$. Alice and Bob time references were synchronized using a classical channel providing a $10$ MHz clock signal. The
detections from the \textit{single-photon detectors} (SAPD) were
recorded by a timetagging unit and then transferred to a computer
to derive the relevant parameters (e.g. \textit{Lower Bound Secure Key Rate}
(LBSKR), \textit{Raw Key Rate} (RKR) and \textit{Quantum Bit Error
Ratio} (QBER)). The parameters of interest were extrapolated using enough accumulated detections in a burst mode operation, which for high RKR it accounted for about $100$ ms while for the lowest RKR it accounted for $16.6$ minutes. The timetagging unit allows $10$ Mcps transmission to the computer, by \textit{direct memory access} (DMA), and also it has a timestamp resolution of $78.125$ ps. Accounting for $500$ ps jitter of SAPD (Perkin Elmer SPCM-AQ4C), $100$ ps jitter and $400$ ps of the width of the generated optical pulses, a $1$ ns time window was implemented by software, allowing to reduce the background noise.

The free-space link was emulated in the laboratory by placing
different neutral density attenuators along the propagation path
from Alice (the source) to Bob (the receiver). We then
measured the rates and performances of the QKD decoy state BB84 protocol using the presented source. In the implementation of
the BB84 protocol, only single-photon pulses contribute to
the secure key, while in the 3-state decoy state protocol the
full range of intensities should be considered.

All but privacy amplification was implemented, were the lower bound for the secure key generation
rate can be obtained as (security proof derived in \cite{Wang2005a,PhysRevLett.94.230504})
\begin{equation}
R\geq q \frac{N_\mu}{t}\left\{-Q_\mu f\left(E_\mu\right)H_2\left(E_\mu\right)+Q_1\left[1-H_2\left(e_1\right)\right]\right\}
\end{equation}
where $q$ depends on the implementation ($1/2$ for the BB84 protocol), $N_{\mu}$ is the total number of signal pulses sent, $t$ is the time duration of the QKD transmission, $\mu$ represents the intensity of the signal states, $Q_\mu$ is the gain of the signal states, $E_\mu$ is the signal pulses QBER, $Q_1$ is the gain of single photon states, $e_1$ is the error rate of single photon states, $f\left(x\right)$ is the bi-direction error correction efficiency (taken as $1.16$ \cite{Brassard1994}, for an error ratio of $1$\%) and $H_2\left(x\right)$ is the binary Shannon information function, given by
\begin{equation}
H_2\left(x\right)=-x\log_2\left(x\right)-\left(1-x\right)\log_2\left(1-x\right)
\end{equation}

\section{Results and discussion}
Figure \ref{Fig:SOA_modulation} shows the SOAs' capability to be driven at $100$ MHz while achieving spatial switching operation and intensity level generation, suitable for decoy state BB84 protocol. In addition low driving voltages are needed, making the design suitable for electronic integration with low electrical power consumption drivers. Furthermore, by design, the phase of each pulse varies at random between pulses due to the fact that, as it was already mentioned, pulses are generated by taking continuously the laser diode above and below threshold, as explained in \cite{Jofre2010JLT-FPS-IMPM}.

The level of indistinguishability is directly reflected on the amount of Secure Key Rate (SKR) that has to be removed in the privacy amplification step, in order to ensure security. The amount of bits to be removed when filtering with interference filters ($10^{-3}$ bits per pulse) can be achieved without using an interference filter but pre-selecting four SOAs with similar low ASE values. Considering the above values for $I(S:B)$, $I(F:B)$ and $I(T:B)$ the corresponding information leakage is small compared to the information leakage indicated by the QBER of the contribution of pulses with more than one photon.

\begin{figure}[htbp]
\centering\includegraphics[width=10cm]{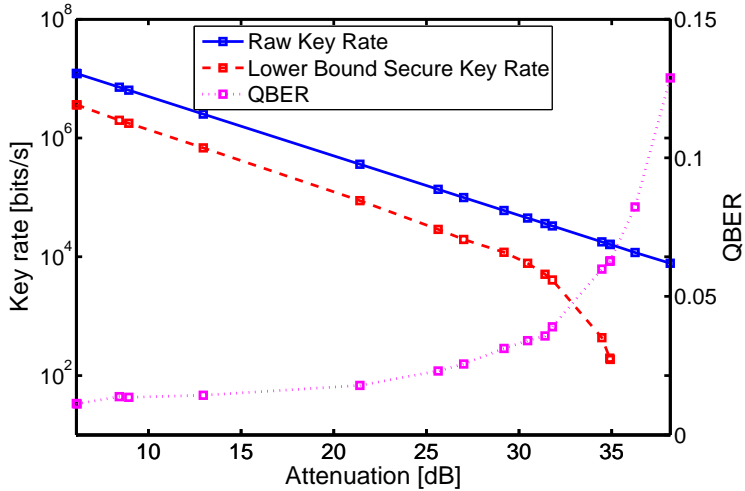}
\caption{QKD BB84 implementing decoy states transmission results. Raw Key Rate (blue solid line), Lower Bound Secure Key Rate (red dashed line) and QBER (magenta dotted line). The detectors' efficiency was set to 50$\%$, background yield $5.58\times 10^{-4}$, $2$ dB were accounted for losses due to the transmitting and receiving optical systems.}
\label{Fig:FPS-4SOA_QKD_measurement_result_attenuation}
\end{figure}

Figure \ref{Fig:FPS-4SOA_QKD_measurement_result_attenuation} shows the detected RKR and QBER, and the resulting LBSKR, for different link attenuations. In the transmission measurement the detectors' efficiency was set to $50\%$, $2$ dB were accounted for losses due to the transmitting and receiving optical setup with a background yield of $5.58\times 10^{-4}$. The background yield $Y_0$ includes the detectors' dark count and other background contributions from stray light \cite{PhysRevA.72.012326} being for larger distances the major cause of secure key rate drop. As expected, for signal rates well above the noise floor the RKR decreases exponentially as the attenuation increases, so does the LBSKR. When the signal rate decreases the noise starts to dominate, so the QBER increases rapidly as well as the LBSKR, until the QBER is $>0.11$ when the LBSKR drops completely to zero. In particular, results for the emulated decoy state BB84 protocol transmission, for a particular attenuation of $6$ dB, are shown in Table \ref{Tab:Relevant_parameters_QKD_emulation}. We have achieved a LBSKR of $3.64$ Mbps with a QBER as low as $1.14\times 10^{-2}$ for an attenuation of $6$ dB while the achieved LBSKR became $187$ bps for an attenuation as high as $35$ dB.
\begin{table}
\centering \caption{Summary of the QKD results for a BB84 transmission, implementing the decoy state protocol. The obtained values are for a $6$ dB attenuation, where $\mu$, $\nu_1$ and $\nu_2$ are the signal, decoy 1 and decoy 2 (ideally vacuum) states. The computed values are the gains for the signal $Q_\mu$, decoy 1 $Q_{\nu_1}$, decoy 2 $Q_{\nu_2}$ states and the QBER for the signal states $e_\mu$. Finally the lower bound of the secure key rate $R_{secure}$, for the presented source, is $3.64$ Mbps with a QBER as low as $1.14\times 10^{-2}$ while a $R_{secure}$ of $187$ bps for an attenuation as high as $35$ dB.}
\label{Tab:Relevant_parameters_QKD_emulation}
%\resizebox{\columnwidth}{!}{
\begin{tabular}{l l c l l}
\multicolumn{5}{c}{}\\
\hline
            Parameter & Value & \vline & Parameter & Value\\
\hline
            Attenuation & $6$ dB & \vline & $Q_\mu$ & $1.18\times 10^{-1}$\\
            $\mu$ & $0.5$ & \vline & $Q_{v_1}$ & $1.8\times 10^{-2}$\\
            $\nu_1$ & $6.6\times 10^{-2}$ & \vline & $Q_{v_2}$ & $3\times 10^{-3}$\\
            $\nu_2$ & $2\times 10^{-3}$ & \vline & $e_\mu$ & $1.14\times 10^{-2}$\\
            $R_{secure}$ & $3.64$ Mbps & \vline & $f\left(E_\mu\right)$ & $1.16$\\
\hline
\end{tabular}%}
\end{table}

\section{Conclusions}\label{Conclusions}
We have shown that a single photon source based on an attenuated laser diode for QKD applications can be built based on a novel scheme including semiconductor optical amplifiers. The source is capable of generating pulses of random polarization distributed over four states and three intensity levels required for decoy state BB84 protocol. A lower bound secure key rate of $3.64$ Mbps with a quantum bit error ratio as low as $1.14\times 10^{-2}$ for an attenuation of $6$ dB. To our knowledge, this is the fastest polarization encoded QKD system which has been reported so far. Given the relatively low driving voltages of the SOAs, the laser diode and the other integrated optical components, the proposed transmitter is potentially low power consumption, highly integrable and stable. The experimental demonstration has been carried out at $850$ nm, for the implementation in free-space links, with $100$ MHz generation rates. However, taking into consideration that the SOA's bandwidth can go well beyond $10$ GHz and operate also at other wavelengths (e.g. $1310$ nm and $1550$ nm for fiber transmission), the source can be easily scalable to higher bit rates, the upper limit probably being set by the laser diode itself.

\section*{Acknowledgments}
This work was carried out with the financial support of the Ministerio de Educación y Ciencia (Spain) through grants TEC2010-14832, FIS2007-60179, FIS2008-01051 and Consolider Ingenio CSD2006-00019, and also by the European Space Agency under ARTES-5 telecom programme Contract No. 21460/08/NL/IA.

\bibliographystyle{IEEEtran}

\end{document}